# Perceptive Statistical Variability Indicators

**Kalman Ziha** University of Zagreb,
Faculty of Mechanical Engineering and Naval Architecture,
Department of Naval Architecture and Ocean Engineering,
Ivana Lucica 5, 10000 Zagreb, Croatia
E-mail: **kziha@fsb.hr**
**Tel: +385 1 6168132**
**Fax: +385 1 61656940**

**Abstract:** The concepts of variability and uncertainty, both epistemic and alleatory, came from experience and coexist with different connotations. Therefore this article attempts to express their relation by analytic means firstly setting sights on their differences and then on their common characteristics. Inspired with the alternative expression of uncertainty defined as the average number of equally probable events based on entropy concept in probability theory, the article introduced two related perceptive statistical measures which indicate the same variability and invariability as the basic probability distribution. First is the equivalent number of a hypothetical distribution with one sure and all the other impossible outcomes which indicates variability. Second is the appropriate equivalent number of a hypothetically uniform distribution with all equal probabilities which indicates invariability. The article interprets the common properties of variability and uncertainty on theoretical distributions and on ocean-wide wind wave directional properties compiled in the Global Wave Statistics.

## 1 Introduction

The variability assessment of $N$ discrete numbers $y_i,\ i=1,2,...,N$, is a problem of lasting interest in statistics and elsewhere (e.g. Cramer, 1945, Kenney and Keeping, 1951, Anderson and Bancroft, 1952, Hogg and Craig, 1965). A set of $N$ normalized numbers $p_i,\ i=1,2,...,N$ can represent a discrete distribution of probabilities **P** (1) in the range ($p_{Min} - p_{Max}$) as shown:

$$\mathbf{P}_N = \left(p_1, p_2, \cdots, p_N\right) \tag{1}$$

The disjoined random events $E_j$ with probabilities of events $p_i = p(E_i),\ i=1,2,\cdots,N$ configure a system $\mathbf{S}_N$ that can be written down after Khinchin (1957) in a form of an $N$-element finite scheme:

$$\mathbf{S}_N = \begin{pmatrix} E_1 & E_2 & \cdots & E_j & \cdots & E_N \\ p_1 = p\left(E_1\right) & p_2 = p\left(E_2\right) & \cdots & p_j = p\left(E_j\right) & \cdots & p_N = p\left(E_N\right) \end{pmatrix} \tag{2}$$

The probability of a distribution of $N$ probabilities $\mathbf{P}_N$ (1) or of a system of $N$ events $\mathbf{S}_N$ (2) is then in general $p_N(\mathbf{P})\, or\, p_N(\mathbf{S}) = \sum_{i=1}^{N} p_i \le 1$. For a complete distribution is $p_N(\mathbf{P})\, or\, p_N(\mathbf{S}) = 1$.

## 2 Variability of probabilities of probability distributions

For generally partial distributions $\mathbf{P}_N$ (1) the mean value of $N$ probabilities is:

$$\overline{p} = \overline{p}(\mathbf{P}_N) = (1/N)\cdot p(\mathbf{P}_N) \tag{3}$$

The average variance $V(\mathbf{P}_N)$ of $N$ probabilities $p$ of a distribution $\mathbf{P}_N$ (1) reads:

$$V_N(\mathbf{P}) = \sigma_N^2(\mathbf{P}) = (1/N)\cdot\sum_{i=1}^{N}(p_i - \overline{p})^2 = (1/N)\cdot\sum_{i=1}^{N} p_i^2 - \overline{p}^2 \tag{4}$$

The article considers next a proposal for a reference value of variance (4) in order to describe a probability distribution when one probability $p_j \to p(\mathbf{P}_N)$ is dominant and all the others $p_{i\neq j} \to 0$ for $i = 1,2,\cdots N-1$ are vanishing. The reference value of variance can be then calculated by definition as:

$$V_{\text{Ref}}(\mathbf{P}_N) = \lim_{\substack{p_j \to p(\mathbf{P}) \\ p_{i\neq j} \to 0}} V(\mathbf{P}_N) = p(\mathbf{P}_N)\cdot(N-1)/N^2 \tag{5}$$

Appendix A presents the proof that the reference variance (5) is the maximally attainable value.

The coefficient of variation of $N$ probabilities of a distribution of probabilities $\mathbf{P}_N$ (1) it reads:

$$CV_N(\mathbf{P}) = \sigma_N(\mathbf{P})/\overline{p} \tag{6}$$

The coefficient of variation of $N$ probabilities (6) has continuity in its arguments, monotonic increase with the number of outcomes up to limiting value $CV_{\mathrm{Ref}}(\mathbf{P}) = \sqrt{N-1}$ and a composition rule based on additivity rule of variances. Impossible occurrences with zero probability do not influence affect variability (6) but the incompleteness of distributions $p_N(\mathbf{P})<1$ affects variability (4). The coefficient of variation of $N$ probabilities (6) of a distribution of probabilities $\mathbf{P}_N$ (1) can be appropriately presented also by its relative value $cv_N(\mathbf{P}) = CV_N(\mathbf{P}) / \sqrt{N-1}$.

For a range of discrete random variables $n$=1, 2, 5, 10 and 50 and for a range of probabilities $p$ from 0 to 1, the variability of binomial distributions is presented by the coefficient of variation (6), (Fig. 1).

## 3 Uncertainty of systems of events

The concept of entropy has been introduced earlier in information theory for assessment of the amount of information pertinent to system of events $\mathbf{S}_N$ (2) (Hartley, 1928, Shannon and Wiever, 1948). The entropy has been lately applied in probability theory to define the uncertainties of systems of events $\mathbf{S}_N$ (2) (Khinchin, 1957, Renyi, 1970, Aczel and Daroczy, 1975). Entropy defines the uncertainty of the observable properties that turns into information when observations become available.

The entropy of a single event is defined as the logarithm of the equivalent number of events $1/p(E_i)$ with equal probability $\mu(E_i) = \log_2\left[1/p(E_i)\right]$ and can be interpreted according to Wiener (1948) either as a measure of the information yielded by the event $E_i$ or how unexpected the event was.

The uncertainty of a complete system $\mathbf{S}_N$ (2) of $N$ events can be expressed as the weighted sum of unexpectednesses of all events by the Shannon's entropy (Shannon and Weaver, 1949):

$$H_N(\mathbf{S}) = \sum_{j=1}^{N} p_j \cdot \mu_j = -\sum_{j=1}^{N} p_j \log p_j \qquad (7)$$

According to Cover and Thomas (2006) the Shannon's information entropy $H$ has a number of natural properties for notions of uncertainty: continuity in its arguments, monotonic increase with the number of equi-probable outcomes and a composition rule. The uncertainty of an incomplete system of $N$ events $\mathbf{S}_N$ (2) can be defined as the limiting case of the Renyi's entropy (1970) of order 1 using (3) and (7) as:

$$H_N^{R1}(\mathbf{S}) = -\frac{1}{p(\mathbf{S})}\sum_{j=1}^{N} p_j \log p_j \qquad (8)$$

Shannon's axiomatic derivation of entropy explains (1949) why it is the intuitive measure of uncertainty. In addition, the uniqueness theorem by Khinchin (1957) proves that the entropy is the only function that measures the probabilistic uncertainty of systems of events in agreement with the experience of uncertainty. The theorem of mixture of distributions (Khinchin, 1957; Renyi, 1970) provides the conditional (average) entropy of system $\mathbf{S}$ with respect t subsystems of events.

The uncertainty of systems of events **S** (2) for binomial distributions for a range of probabilities $p$ from 0 to 1 and for numbers of variables $n$=1, 2, 5, 10, and 50 is presented by the entropy (7) (Fig. 1).

**5 Average numbers of events and equivalent numbers of outcomes**

Aczel and Daroczy mentioned earlier (1975) the average number of equally probable events that follows from the condition of maximal uncertainty $H_N(\mathbf{S}) = \log_2 F_N(\mathbf{S})$ as an uncertainty indicator:

$$F_N(\mathbf{S}) = 2^{H_N(\mathbf{S})} \tag{9}$$

The average number of events $F_N(\mathbf{S})$ (9) in the range $1 \le F_N(\mathbf{S}) \le N$ is not any more dependent on the base of applied logarithm. It defines a hypothetically uniform probability distribution of $F_N$ equal probabilities amounting to $1/F_N$ with same uncertainty as the entropy $H_N(\mathbf{S})$ of the system **S** (Fig. 2). Relative measures such as $h_N(\mathbf{S}) = H_N(\mathbf{S})/\log N$ (7, 8) or (9) $f_N(\mathbf{S}) = F_N(\mathbf{S})/N$ can be appropriate.

The concept of average numbers of events based on entropy (9) has inspired the investigation of equivalent numbers of outcomes based on statistical variability of probability distributions (3-6). The article firstly concentrates on variability of probability distributions. Therefore it defines the equivalent number $G_N(\mathbf{P})$ of a hypothetical probability distribution with one sure and remaining $G_N(\mathbf{P})$-1 impossible outcomes which provides the same variability as the basic distribution **P** of $N$ probabilities (Fig. 2). The equivalent number $G_N(\mathbf{P})$ follows directly from the condition that the coefficient of variation $CV_N(\mathbf{P})$ (6) is equal to its maximal value $\sqrt{G_N(\mathbf{P}) - 1}$, as it is shown below:

$$G_N(\mathbf{P}) = CV_N^2(\mathbf{P}) + 1 \tag{10}$$

In the context of this research the square $CV_N^2(\mathbf{P})$ of the coefficient of variation (6) of any probability distribution **P** (1) or a system of events **S** (2) in (10) represents the number of impossible outcomes or events in addition to a single sure outcome or certain event that provide the same variability or certainty as the original probability distribution of N outcomes or the system of $N$ events.

The article next focuses on invariability of probability distributions. The equivalent number $D_N(\mathbf{P})$ of equal probabilities in amount of $1/D_N(\mathbf{P})$ of a hypothetically complete probability distribution expresses the invariability when the variance (4) of a distribution **P** (1) of $N$ probabilities is equal to zero $\left[1/D_N(\mathbf{P})\right]\cdot\sum_{i=1}^{N} p_i^2 - 1/D_N^2(\mathbf{P}) = 0$ (Fig. 2), as it is shown below:

$$D_N(\mathbf{P}) = 1/\sum_{i=1}^{N} p_i^{\,2} = N/\left[p_N^2(\mathbf{P})\cdot\left(CV_N^2(\mathbf{P})+1\right)\right] \tag{11}$$

In the context of this research the sum of squares of probabilities $\sum_{i=1}^{N} p_i^{\,2}$ in (4) of any probability distribution **P** (1) or a system of events **S** (2) in (11) represents the mean probability of a hypothetically uniform probability distribution or a system of events of $D_N(\mathbf{P})$ equally probable outcomes or events. The following relation links the two equivalent numbers for generally incomplete distributions with known probability $p_N(\mathbf{P})$ of $N$ possible outcomes:

$$D_N(\mathbf{P})\cdot G_N(\mathbf{P}) = N/p_N^2(\mathbf{P}) \tag{12}$$

The relation (12) in logarithmic form expresses uncertainties as shown:

$$\log D_N(\mathbf{P}) + \log G_N(\mathbf{P}) = \log N - 2logp_N(\mathbf{P}) \tag{13}$$

The terms (9), (10) and (11) imply the generalization of number of events other than integers.

The increasing number of impossible outcomes in the range $0 \le G_N(\mathbf{P}) - 1 \le N - 1$ (10) with respect to one sure outcome based on $CV_N(\mathbf{P})$ (6) indicates increasing variability and rise of predictability. On the other hand, the increasing value of equivalent number of equally probable outcomes (11) in the range $1 \le D_N(\mathbf{P}) \le N$ (11) indicates lessening variability and a drop of predictability (Table 1).

Simultaneously, the average number of equally probable events (9) in the range $1 \le F_N(\mathbf{S}) \le N$ based on entropy (7, 8) represents rise in uncertainty and as such indicates drop of predictability (Table 1). Thus, the equivalent numbers of outcomes $D_N(\mathbf{P})$ (11) and the average numbers of events $F_N(\mathbf{P})$ (9) go well together in expressing the invariability-uncertainty-unpredictability (Table 1).

The increasing equivalent probability in the range $1/N \le 1/D_N(\mathbf{P}) \le 1$ based on statistical invariability (11) expressing the growth of invariability and the average probability $1/N \le 1/F_N(\mathbf{S}) \le 1$ based on probabilistic entropy (7, 8) representing the drop of uncertainty, indicate rise of predictability. Hence, he equivalent numbers of events $G_N(\mathbf{P})$ (9) go well together with equivalent $1/D_N(\mathbf{P})$ and average $1/F_N(\mathbf{S})$ probabilities in expressing the variability-certainty-predictability (Table 1).

The difference between analytical definitions of variability (10) and uncertainty (9) is in perception of impossible or certain events. The uncertainty (9) vanishes whenever there is at least one certain event regardless of the number of impossible events. However, when any one event is sure, the variability perception (10) depends on the number of remaining impossible events since the mean value of the distribution of probabilities changes with overall number of outcomes.

Statistical properties (3-6) of distributions of probabilities $\mathbf{P}_N$ (1), the entropy (7, 8) of systems of events $\mathbf{S}_N$ (2) or the average and equivalent numbers (9, 10, 11) do not depend on sequences of probabilities. Variability and uncertainty are commonly considered as objective properties since they depend on nothing else but on the probability distributions. Therefore the intrinsic predictability based on statistical variability can be considered as an objective property too. However, probabilistic forecasts can be performed with conditional distributions employing posterior distributions. The common posterior verification method for probabilistic forecasts is the Brier score (1950) proposed as the average deviation between predicted probabilities for a set of events and their outcomes. Relative measures of predictability can be also important given that a Bayesian viewpoint of prediction is a useful one. In that context the relative entropy is useful and worth considering, e.g. Kleeman (2002), Roulston&Smith (2002) and Bröcker (2009). This of course involves both prior and posterior distributions.

## 6 Variability and uncertainty of wind wave climate

Visual observations of wind speeds (Beaufort Scale) and directions and wave heights have been reported since 1854; observations of commercial ships have been archived since 1861; wave height, period and directions have been reported from ships in normal service all over the world since 1949. The observations are systematically collected following the non-instrumental methodology prescribed since 1961 by the resolution of the World Meteorological Organization (WMO) in order to assure that the data are globally homogeneous in quality and covering most sea areas of practical interest for shipping, navigation, towing and offshore activities. The compilation of these observations for each of the $N_A$=104 Marsden's squares (Fig. 4) is the Global Wave Statistics (GWS) prepared by Hogben, Dacunha & Olliver (1986) that uses the past experiences to eliminate biases.

The study in the sequel investigates the variability of wave directions based on annual wind wave climate observation reported in the GWS (1986) by probability distributions in 104 ocean areas *A* (Fig. 4 and 6) as $\mathbf{P}_8(A) = (\mathrm{d}_N,\ \mathrm{d}_{NE},\ \mathrm{d}_E,\ \mathrm{d}_{SE},\ \mathrm{d}_S,\ \mathrm{d}_{SW},\ \mathrm{d}_W,\ \mathrm{d}_{NW})$ of *N*=8 principal wave directions.

For wind wave climate directional observations circular statistics is appropriate, e.g. Fisher (1993). However, the variability of probabilities of wind wave directions is not necessarily of circular character and linear descriptive statistics can be applied.

The article firstly graphically presents the equivalent numbers of directions $D_8$[**P**(A)] (11), the average number of events $F_8$[**P**(A)] (9) with respect to the nominal number of directions *N*=8 for all ocean areas (Fig. 3). The two numbers indicate same ordering based on variability and uncertainty considerations. The same graph also presents the relative uncertainty $h_8$[**P**(A)] (8),and the statistical variability of probabilities of wave directions $cv_8$[**P**(A)] (6) which indicate opposite ordering (Fig. 3).

The article next presents the chart of the relative probabilistic statistical variability $cv_8$[**P**(A)] (6) (Fig. 4) and the chart of equivalent numbers of events $D_8$[**P**(A)] (11) (Fig. 5) of wind wave directions.

There follows few comments. The wave directions are highly predictable in some areas in the eastern Pacific Ocean such as A64 given by distribution $\mathbf{P}_8$**(**A64)=(0.0042  0.0098  0.1151  0.6081  0.2110 0.0234  0.0049  0.0033) where $cv_8$(A64) ~60%] (Fig. 4). The three directions (east, south-east and south) prevail with about 90% (Fig. 6). The appropriate equivalent number of wave directions $D_8$(A64)=2.3 (Table 2 and Fig. 5) indicates invariability of directions equivalent to probability distributions with numbers of equally probable outcomes between two $\mathbf{P}_8$(A64)=(1/2  1/2) and three $\mathbf{P}_8$(A64)=(1/3  1/3 1/3), closer to two than to three. The average number of wave directions $F_8$(A64)=3 (Table 2) indicates uncertainty appropriate to probability distributions with three equally probable outcomes.

The appropriate equivalent number of wave directions $G_8$(A64)=3.6 (Table 2 and Fig. 6) indicates variability equivalent to a probability distributions with number of outcomes from which one is sure and two $\mathbf{P}_8$(A64)=(1 0 0) or three $\mathbf{P}_8$(A64)=(1 0 0 0) are impossible. Similar is the situation in some areas of western Atlantic Ocean such as A66, A67 and A68.

In some ocean areas the wave directions are unpredictable since the directions are almost uniformly distributed. For example in South Pacific area A86 the probability distribution of wave directions is $\mathbf{P}_8$(A86)=(0.1192 0.0941 0.1157 0.1125 0.1299 0.1370 0.1489 0.1152) and the relative coefficient of variation is only $cv_8$(A86)=4.75%. There, the equivalent number of wave directions $D_8$(A86)=8.3 and the average number of wave directions $F_8$(A86)=8.2 (Table 2 and Fig. 7) even exceeds the nominal value of $N$=8 due to incompleteness of observations. This indicates that the equivalent probability distribution is $\mathbf{P}_8$(A86)=(1/8 1/8 1/8 1/8 1/8 1/8 1/8 1/8).

The appropriate equivalent number of wave directions $G_8$(A86)=1.01 (Table 2 and Fig. 7) indicates almost maximal variability, that is almost full certainty, when one outcome is sure and another is close to be impossible $G_8$(A86)=(1 0). Similar is the situation in North Atlantic area A1 where $cv_8$(A1)=4.75%.

## 7 Conclusions

Variability and uncertainty are recognizable as two opposite properties which naturally motivate conscious observers of random phenomena for predictions. Therefore the article advocated two particular functionals using statistical variability defined on probability sets to bring these two properties closer to common experience of randomness. The two types of equivalent numbers of outcomes or events were introduced with the aim to represent common indicators of invariability and certainty as well as variability and uncertainty, other than just statistical variance and probabilistic entropy. The properties of proposed indicators were premeditated to match human comprehension of random phenomena closely to everyone’s gambling perception of hazardous games. For example, it is intuitively perceptive that the flipping of a balanced coin is as predictable as 2 events with probabilities 1/2 and tossing of an unbiased dice as 6 events with probabilities 1/6. The equivalent and average numbers imply the analytical generalization of numbers of outcomes of probability distributions or of numbers of events of systems of events for perceptive presentation o variability and uncertainty other than only integers.

## Appendix A

Let's consider a complete probability distribution of *N* probabilities $\sum_{i=1}^{N} p_i = 1$.

From the Jensen inequality directly follows the lower limit of the sum of squares of a probability distribution $\frac{1}{N} \leq \sum_{i=1}^{N} p_i^2$ . According to common reasoning for any $p_i < 1$ is $p_i^2 < p_i$, and therefore the upper limit is $\sum_{i=1}^{N} p_i^2 < 1$. Since only the unity has the property $1^2 = 1$, the sum of squares attains its maximal value in amount of $\sum_{i=1}^{N} p^2{}_i = 1$ only if any of the probabilities is equal to unity $p_i = 1$ and all the other (*N*-1) probabilities are zero $p_{j \neq i} = 0$ .

**Table 1.** Comparative properties of (in)variability and (un)certainty related to predictability

| | |
|---|---|
| **Variability** (Fig. 1)<br>Coefficient of variation of probabilities (6)<br>$CV_N(\mathbf{P}) = \sqrt{\left[N / p_N^2(\mathbf{P})\right] \cdot \sum_{i=1}^{N} p_i^2 - 1}$<br>Min: 0 – Invariable<br>*(All N outcomes equally probable)*<br>Max; $\sqrt{N-1}$ – Maximal variability<br>*(One sure outcome all N-1 others impossible)*<br>Unit: 1– *(One sure outcome, one impossible N=2)* | **Uncertainty** (Fig. 1)<br>The entropy of system of events (8)<br>$H(\mathbf{S}) = -\left[1/ p_N(\mathbf{S})\right] \sum_{i=1}^{N} p_i \log p_i$<br>Min: 0 – Certain<br>*(One certain event, N-1 others impossible)*<br>Max; $\log N$ - Full uncertainty<br>*(All N events equally probable)*<br>Unit: 1 bit *(Two equally probable events* |
| **Invariability** (Fig. 2)<br>Equivalent number of outcomes (11)<br>$D_N(\mathbf{P}) = 1/\sum_{i=1}^{N} p_i^2 = N/\left[p_N^2(\mathbf{P}) \cdot G_N(\mathbf{P})\right]$<br>Range: $1 \le D_N(\mathbf{P}) \le N$ where $G_N(\mathbf{P}) = CV_N^2(\mathbf{P}) + 1$<br>Min: 1: Fully predictable/Certain<br>*One sure outcome, another impossible(N=2) CV=1*<br>*One sure outcome, N-1 impossible* $CV = \sqrt{N-1}$<br>Max: ***N*** – Unpredictable/Uncertain *CV*=0<br>*(All N outcomes are equally probable)*<br>Ref: 2: *(Two equally probable events N=2)* | **Certainty** (Fig. 2)<br>Average number of events (9)<br>$F_N(\mathbf{S}) = 2^{-\sum_{i=1}^{N} p_i \log p_i} = 2^{H_N(\mathbf{S})}$<br>Range: $1 \le F_N(\mathbf{S}) \le N$<br>Min: 1: Certain/Fully predictable<br>*One certain event, one impossible(N=2) H*=0<br>*One certain event, N-1 impossible H*=0<br>Max: ***N***–Uncertain/Unpredictable *H*=log*N*<br>*(All N events are equally probable)*<br>Ref: 2: *(Two equally probable events N=2)* |

**Table 2**. Variability and uncertainty of wave directions in GWS areas A86 and A64

| PP | $p_8(\mathbf{P})$ | $p_{mean}$ (3) | $H_8(\mathbf{P})$ (7) | $F_8(\mathbf{P})$ (9) | $CV_8(\mathbf{P})$ (6) | $cv_8(\mathbf{P})$ | $D_8(\mathbf{P})$ (11) | $G_8(\mathbf{P})$ (10) |
|---|---|---|---|---|---|---|---|---|
| **Range** | **≤1** | **0.1250** | **0-3** | **1-8** | **0-** | **0-1** | **1-8** | **1-8** |
| $\mathbf{P}_{A86}$ | **0.9725*** | **0.1216** | **3.03** | **8.16** | **0.126** | **0.04** | **8.32** | **1.02** |
| $\mathbf{P}_{A64}$ | **0.9798*** | **0.1225** | **1.59** | **3.01** | **1.57** | **0.59** | **2.33** | **3.57** |

*The considered wave direction observations represent incomplete distributions

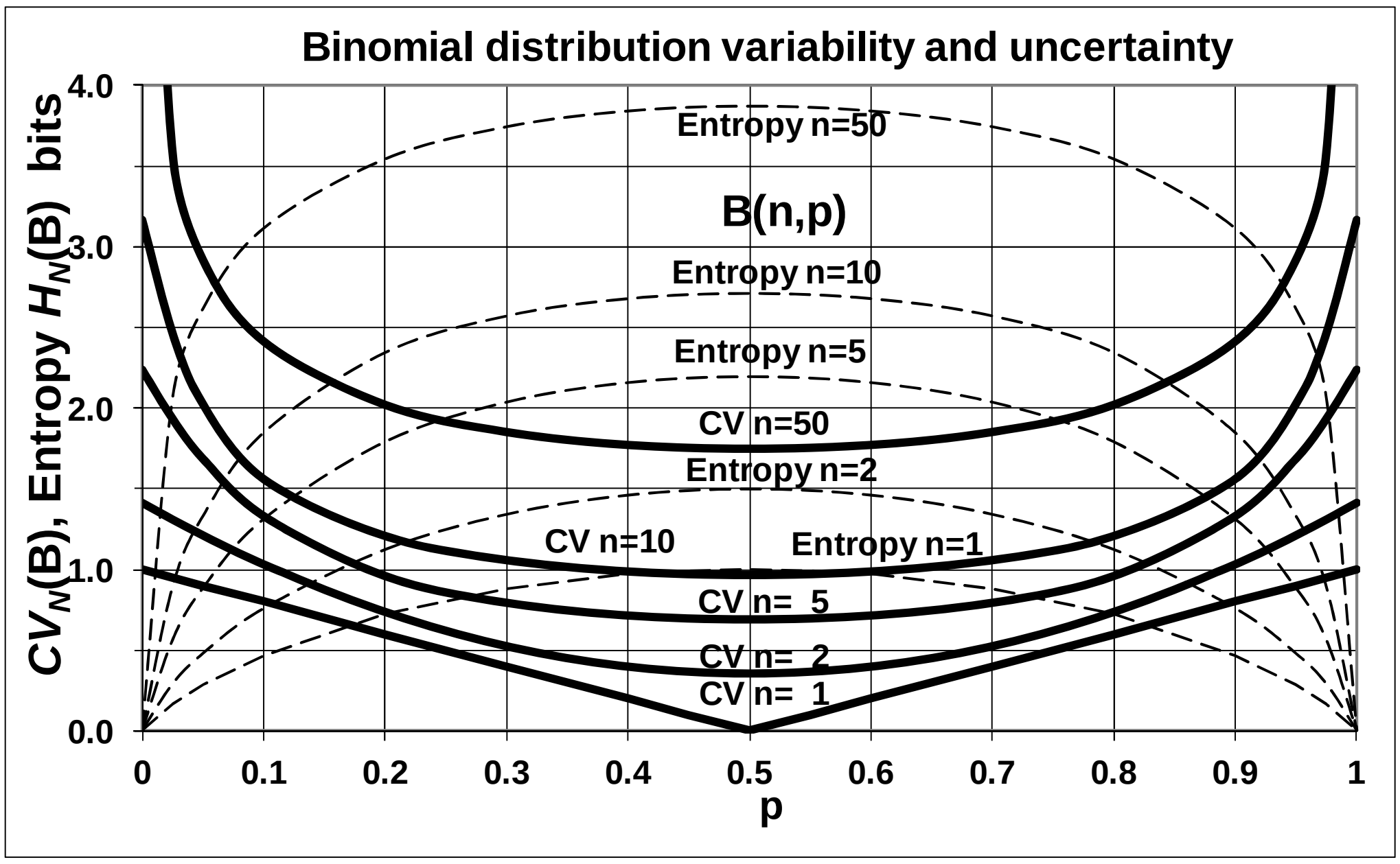

**Figure 1.** Coefficient of variation and entropy of probabilities of Binomial distribution **B**(*n*,*p*)

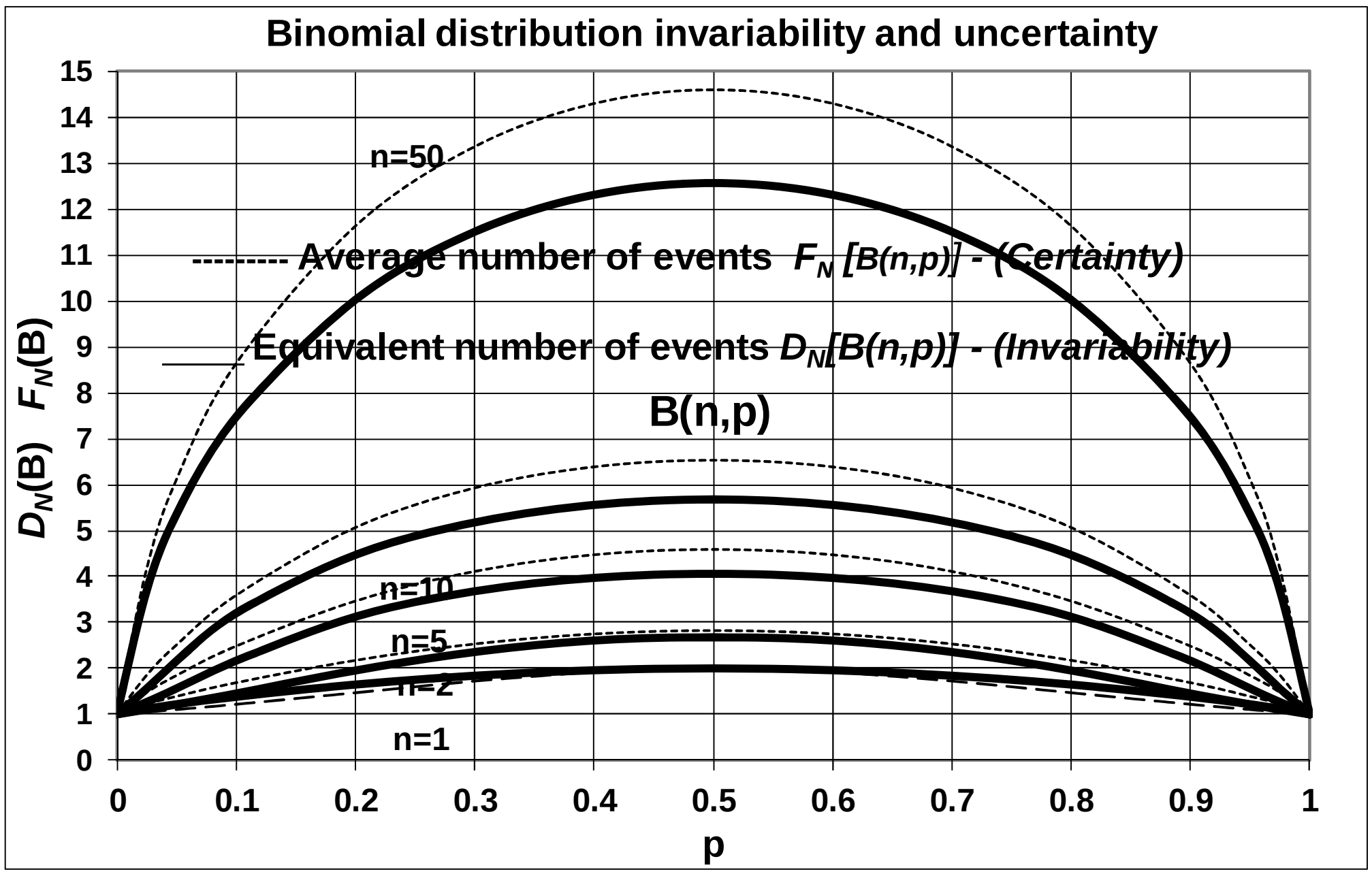

**Figure 2**. Equivalent $G(10)$, $D(11)$ and average $F(9)$ numbers for Binomial distribution $\mathbf{B}(n,p)$

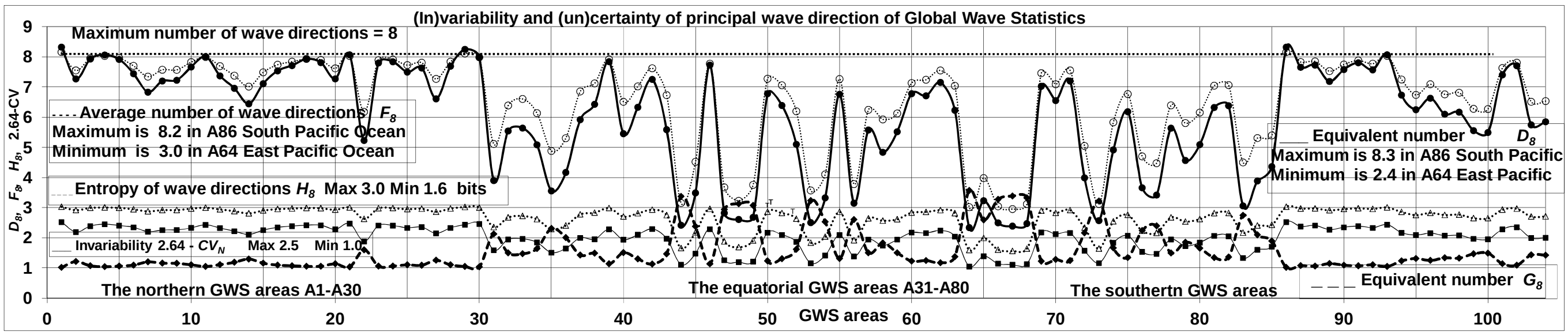

**Figure 3**. (In)variability and (ucertainty of wind wave directions in GWS

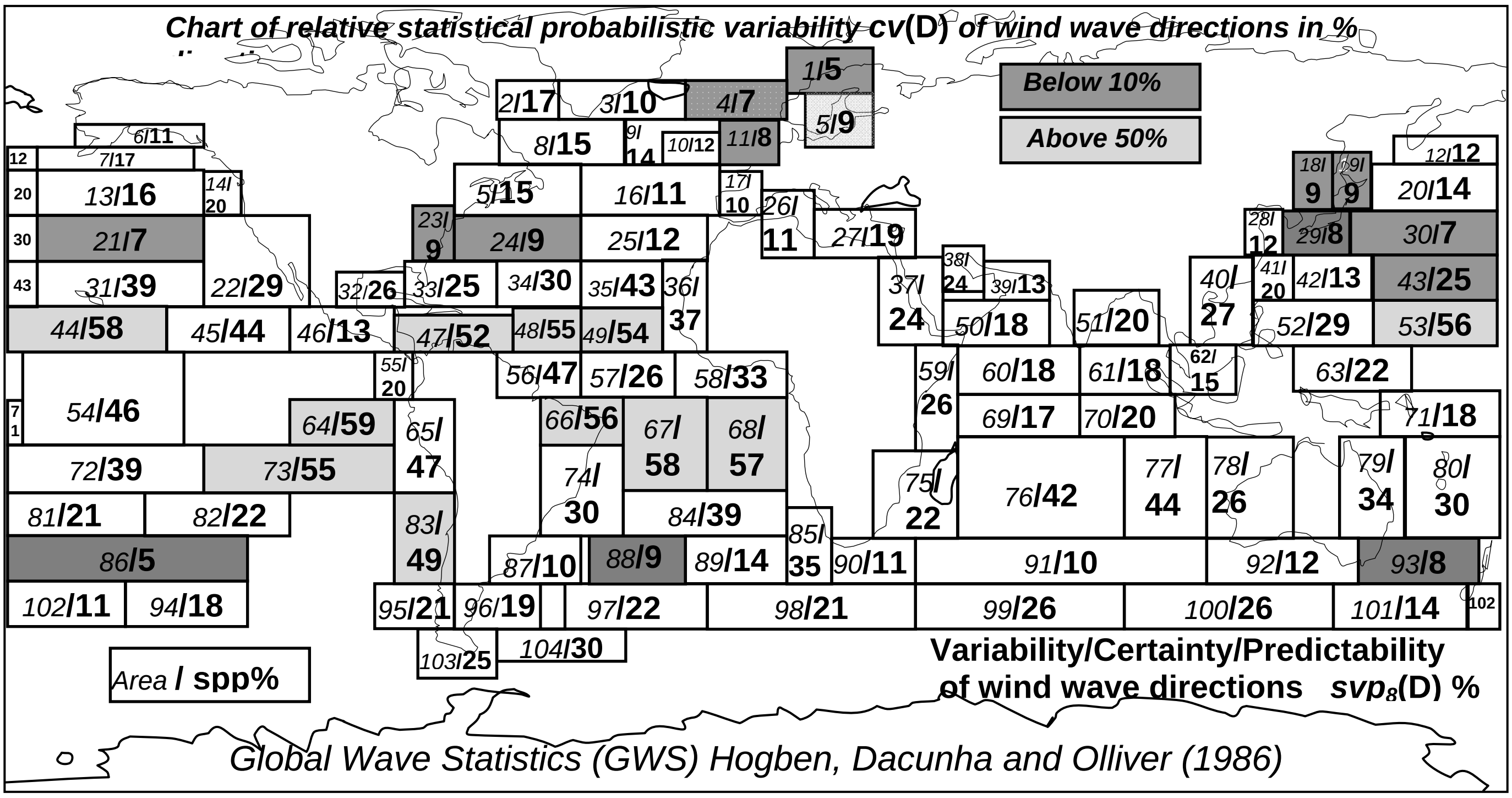

**Figure 4**. Chart of relative statistical probabilistic variability (6) presenting inherent predictability based on probability dosstributions og wind wave directions using prior observations compiled in GWS in %

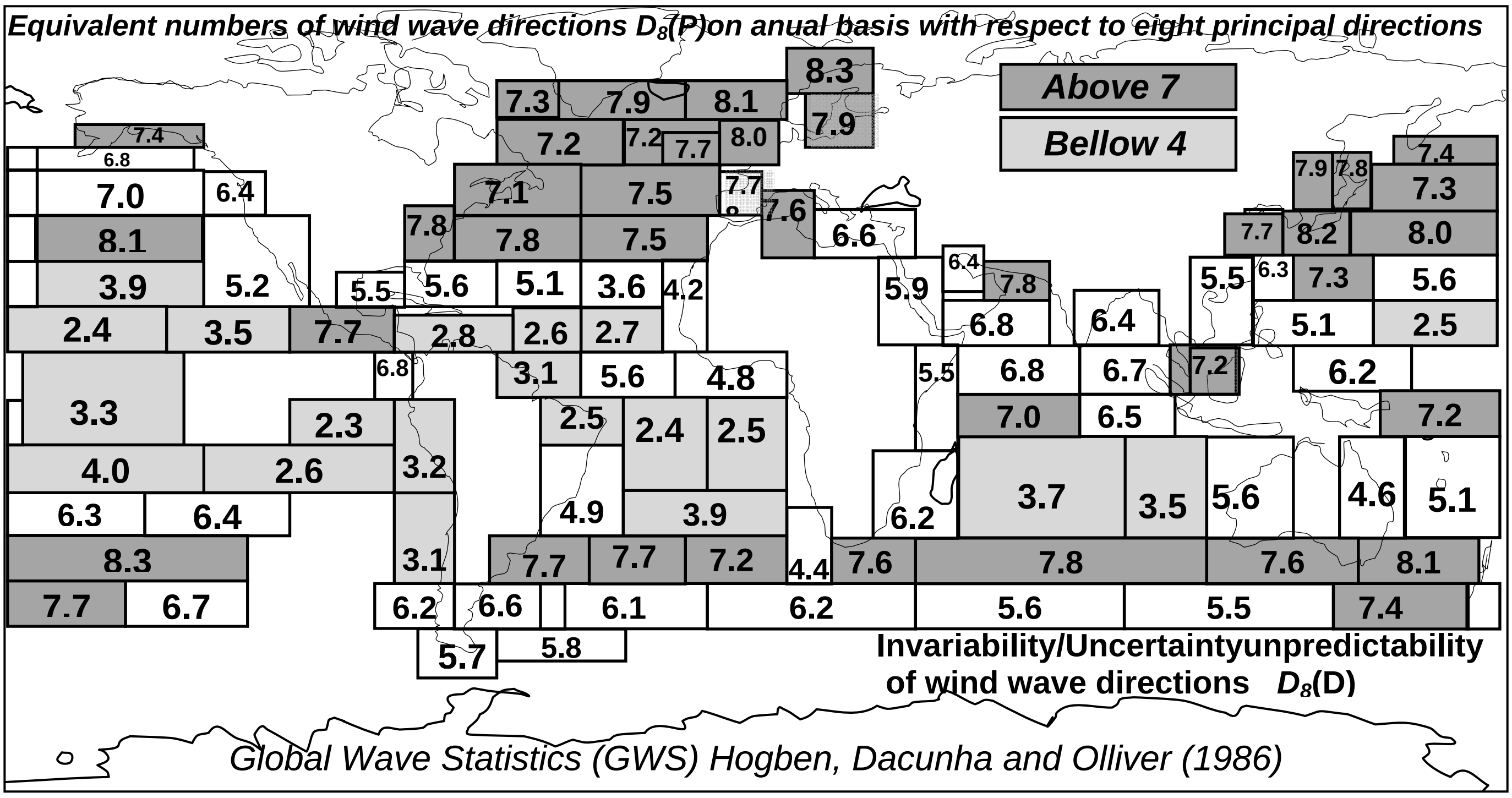

**Figure 5.** Chart of equivalent numbers of equally probable wave directions (11) observed in GWS on annual basis with respect to eight principal directions

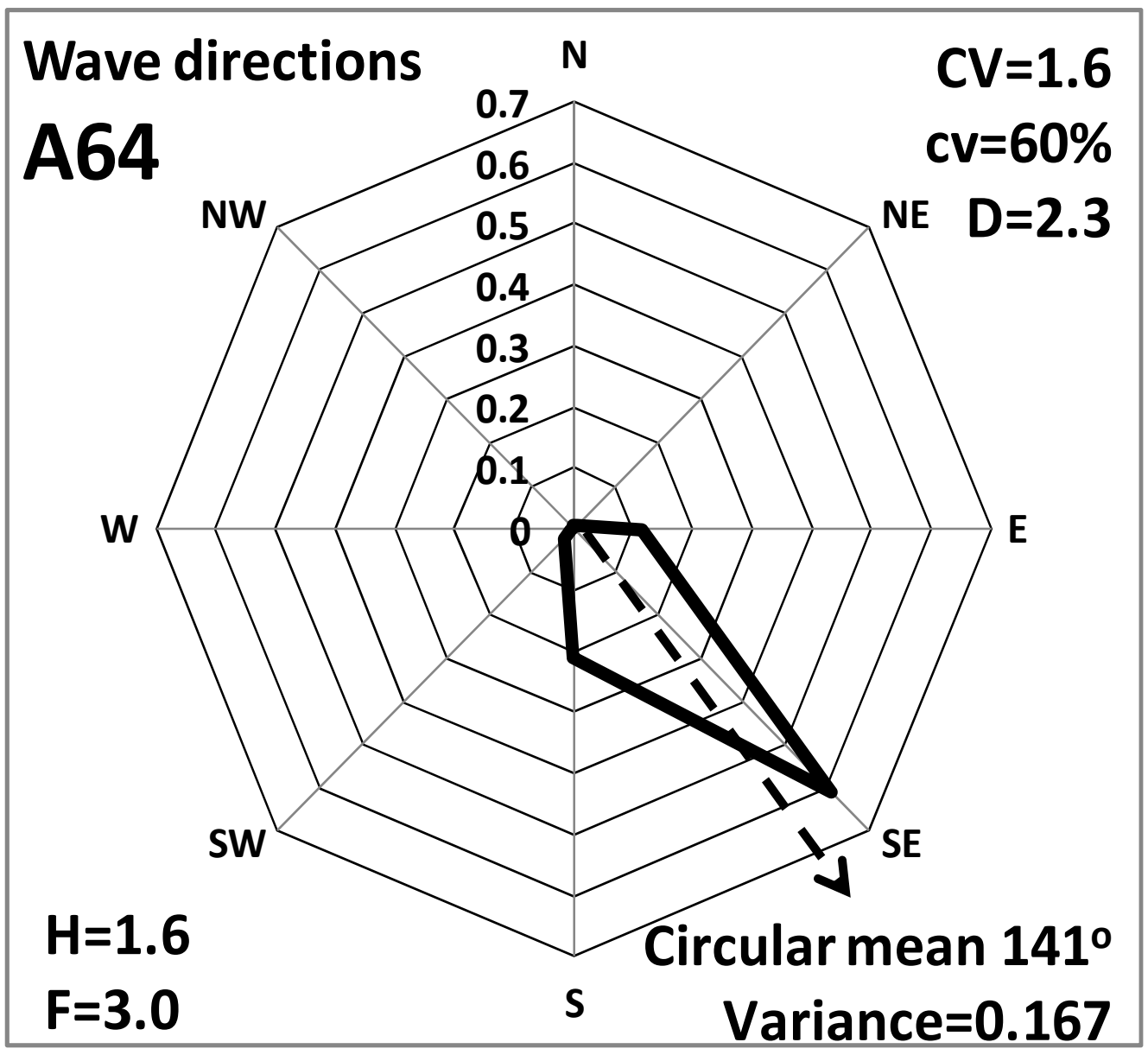

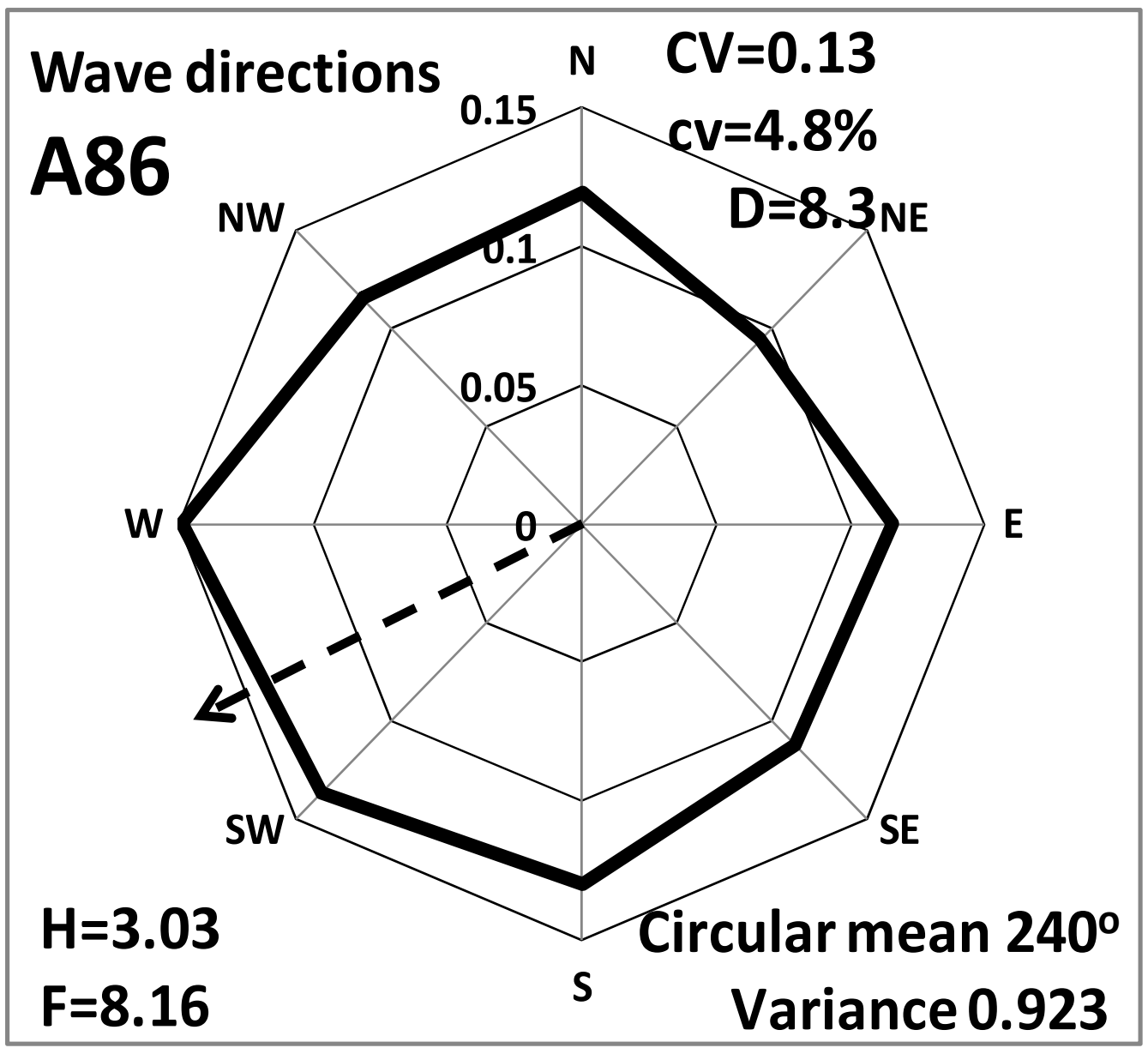

**Figure 6.** Circular distribution of wave directions in ocean area A86 in Nortn Atlantic Ocean

**Figure 7**. Circular distribution of wave directions in ocean area A64 in East Pacific Ocean